\documentclass{article}
\usepackage{fullpage}

\usepackage{xcolor}
\usepackage{tikz}
\usepackage{subcaption}
\usepackage{mathtools}

\usepackage{graphicx,amssymb,amsmath,amsthm}
\usepackage{lineno} 
\usepackage{enumitem}

\usepackage{hyperref}

\begin{document}

\title{The Euclidean $k$-Matching Problem is NP-hard}

\author{ José-Miguel Díaz-Báñez\thanks{Department of Applied Mathematics, University of Seville, SPAIN. \texttt{dbanez@us.es}} \and 
Ruy Fabila-Monroy\thanks{Departamento de Matem\'aticas, Cinvestav, MEXICO. \texttt{ruyfabila@math.cinvestav.edu.mx}}. \and 
José-Manuel Higes-López\footnotemark[1] \thanks{ \texttt{jhiges@us.es}} \and 
Nestaly Marín\thanks{Centro de Investigaci\'on Cient\'ifica y de Educaci\'on Superior de Ensenada, Baja California, MEXICO. {\tt nestaly@cicese.mx}} \and 
Miguel-Angel Pérez-Cutiño\footnotemark[1] \thanks{\texttt{migpercut@alum.us.es}} \and
Pablo Pérez-Lantero\thanks{Universidad de Santiago de Chile (USACH), Facultad de Ciencia, Departamento de Matemática y Ciencia de la Computación, Chile. \texttt{pablo.perez.l@usach.cl.}}
}

\maketitle

%\linenumbers

\begin{abstract}
Let $G$ be a complete edge-weighted graph on $n$ vertices. To each subset of vertices of $G$ assign the cost of the minimum spanning tree of the subset as its weight. Suppose that $n$ is a multiple of some fixed positive integer $k$. The $k$-matching problem is the problem of finding
a partition of the vertices of $G$ into $k$-sets, that minimizes the sum of the weights of the $k$-sets. 
The case $k=3$ has been shown to be NP-hard [Johnsson et al., 1998]. In the Euclidean version, the vertices of $G$ are points in the plane
and the weight of an edge is the Euclidean distance between its endpoints. We call this problem the Euclidean $k$-matching problem.
In this paper we show that, for every fixed $k \ge 3$, the Euclidean $k$-matching is NP-hard through a reduction from Cubic Planar Monotone 1-in-3 SAT. This resolves an open problem in the literature and provides the first theoretical justification for the use of known heuristic methods in the case $k=3$. We also show that the problem remains NP-hard if the trees are required to be paths.
\end{abstract}


\section{Introduction}

Let $G$ be a complete edge-weighted graph on $n$ vertices.
We define the weight of any vertex subset $T\subseteq V$ to be $w(T)$, the cost of its minimum spanning tree. Assume $n$ is divisible by a fixed positive integer $k$.
%To each subset $T$ of the vertices of $G$ assign the cost $w(T)$ of its minimum spanning tree as its weight. Suppose that $n$ is a multiple of some fixed positive integer $k$. 
The \emph{$k$-matching problem} is the problem of finding a partition of the vertices of $G$ into $k$-sets, that minimizes the sum of the weights of the $k$-sets. For $k=2$, this is the problem of finding a minimum-weight perfect matching on $G$. This can be solved in $O(n^3)$ time, e.g. Gabow's~\cite{gabow} implementation of Edmonds' blossom algorithm~\cite{edmonds_1,edmonds_2}. Recent algorithms for weighted matching problems can be found in \cite{gabow2021algorithms}.

Suppose now that the vertices of $G$ are points in the plane and the weight of an edge is equal to the Euclidean distance between its endpoints. We call this instance of the $k$-matching problem, the \emph{Euclidean $k$-Matching Problem (E$k$-MP)}. For the case of $k=2$, geometry can be used to improve on the $O(n^3)$ time algorithm. In 1988, Vaidya~\cite{geometry_helps} gave an $O(n^{2.5} (\log n)^4)$-time algorithm for E2-MP. This was improved by Varadarajan~\cite{geometry_helps2} in 1998 to  $O(n^{1.5} (\log n)^5)$. 
The general 3-matching problem was shown to be NP-complete by Johnsson, Magyar and Nevalainen~\cite{johnsson1998euclidean}; however, the complexity of the specialized Euclidean version of the 3-Matching Problem remained an open problem until now.
The E$k$-MP is phrased as a decision problem as follows:

\vspace{.25cm}

\textbf{The Euclidean $k$-Matching Problem (E$k$-MP) }

\emph{Let $n$ be a multiple of $k$,  $S$ a set of $n$ points in the plane, and $w$ be a weight function such that, for any subset $T\subset S$, $w(T)$ is cost of the minimum spanning tree on the complete graph defined by points in $T$. Given
a parameter $W >0$, does there exist a partition of $S$ into disjoint $k$-sets $\{T_1,T_2,\dots,T_{n/k}\}$ such that \[\sum_{i=1}^{n/k} w(T_i) \le W?\] }

\vspace{.25cm}

Applications of E3-MP and 3-MP include: task assignment (assigning groups of three workers or machines to optimize efficiency), clustering and data analysis (grouping data points into sets of three based on similarity), network design (creating efficient tripartite structures in networks), and logistics and routing (assigning delivery routes in clusters of three locations). Descriptions of such real-world applications can be found in the literature~\cite{crama2012production,johnsson1996determining,magyar2000adaptive}. 

A proof of the NP-hardness of the Euclidean version provides a strong justification for the use of heuristics and approximation algorithms in these applications.
In~\cite{johnsson1998euclidean} the authors provided six lower bounds for the solution of the E$3$-MP and presented experimental results on several heuristics. The same authors later presented 
an exact solution for this problem with empirical performance tests~\cite{magyar1999exact}. Castro Campos et al.~\cite{castro2017integer} introduced three new integer programming models and presented specialized heuristics for the problem, comparing their solutions and execution times against exact models.

In this paper, we prove that the E$k$-MP is NP-hard for $k\geq 3$ using a reduction from a specific version of the 3-SAT problem.
Note that while we prove NP-hardness, establishing full NP-completeness would require showing that solution verification can be done in polynomial time. Since we are dealing with Euclidean distances, this is related to the \emph{Sum of Square Roots Problem} (SSRP)\footnote{The SSRP asks whether, given positive integers $a_1,\dots,a_n$ and an integer $t$, the inequality $\sum_{i=1}^n \sqrt{a_i} \le t$ holds. The difficulty arises in that the square roots might have to be evaluated to a very large precision.},
the complexity of which is an open problem.
The paper is organized as follows.
For expositional clarity, in Section \ref{sec:3sat}, we first introduce the version of $3$-SAT used in our reductions. Then, we prove the NP-hardness of Euclidean $3$-Matching Problem (E$3$-MP) in Section~\ref{sec:3m}, isolating the core geometric gadgets. In Section~\ref{sec:km}, we generalize these ideas to all $k>3$ via a non-trivial extension that handles interdependencies between $k$ points. This layered approach illuminates the evolving complexity of the problem. Finally, in Section~\ref{sec:kpaths}, we extend our hardness results by showing that the E$k$-MP problem remains NP-hard even under the additional constraint that the trees must be paths.

\section{Cubic Planar Monotone 1-in-3 SAT}\label{sec:3sat}

In our reduction, we will use a particular variant of the SAT problem known as Cubic Planar Monotone 1-in-3 SAT.
Let $\psi$ be an instance of $3$-SAT.
If  every variable of $\psi$ appears in exactly $3$ clauses, we say that $\psi$ is \emph{cubic}. Let $G_\psi$ be the bipartite graph where on one side of the partition we have a vertex for each variable of $\psi$ and on the other side we have a vertex for each clause; two vertices are adjacent in $G_\psi$ if and only if the variable vertex appears in the clause vertex. We say that $\psi$ is \emph{planar} if $G_\psi$ is planar. 
We say that $\psi$ is \emph{monotone} if each clause contains either only positive literals or only negative literals (we call these \emph{positive clauses} and \emph{negative clauses}, respectively).
Finally, unlike standard 3-SAT which requires at least one literal to be true in each clause, 1-in-3-SAT is the problem of finding an assignment of truth values to the variables of $\psi$ so that in every clause exactly one of its three literals is satisfied. 
Moore and Robson showed that cubic planar monotone $1$-in-$3$-SAT is NP-complete~\cite{hard_tiling}. In what follows, assume that $\psi$ is an instance of cubic planar monotone $1$-in-$3$-SAT, and that $\psi$ has $n$ variables $x_1,\dots,x_n$.

\begin{figure}
 	\centering
 	\includegraphics[width=0.5\textwidth]{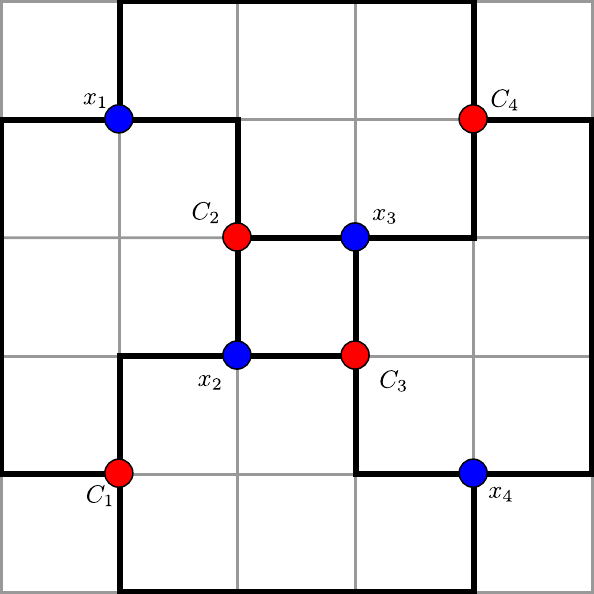}
 	\caption{The graph $G_\psi$, and a grid embedding $D_\psi$, with respect to the formula
 	$\psi:=\protect \overbrace{(x_1 \lor x_2 \lor x_4)}^{C_1} \land (  \protect \overbrace{\overline{x_1} \lor \overline{x_2} \lor \overline{x_3})}^{C_2} \land
 	\protect \overbrace{(x_2 \lor x_3 \lor x_4)}^{C_3} \land \protect \overbrace{(\overline{x_1} \lor \overline{x_3} \lor \overline{x_4})}^{C_4}$ }\label{fig:grid_embedding}
 \end{figure}

\subsection*{Grid Embeddings}

A \emph{planar embedding} of a graph $G$ is a drawing of the graph in the plane where: vertices are mapped to
distinct points; edges are drawn as simple curves joining their respective endpoints and no two edges intersect in their interior. A planar embedding of $G$
defines a cyclic order on the neighbours of every one of its vertices, by considering the clockwise order of the edges
as they appear around each vertex. These orders of the neighbours of each vertex are known as a \emph{rotation system}; it is used to represent 
the embedding since it is invariant under homeomorphisms of the plane. 
An alternative but equivalent way to represent planar embeddings is to give, for each face of the embedding, the circular sequence of edges encountered when walking clockwise around the boundary of that face. This is called a \emph{planar representation}. It is straightforward to obtain a planar representation from a rotation system, and vice versa.
Chiba, Nishizeki and Abe~\cite{linear_embedding} gave a linear-time algorithm that given a planar graph produces the rotation system of a planar embedding of the graph. For our reduction, we need to work with a planar embedding of the graph $G_\psi$ associated with the SAT instance.

A \emph{grid embedding} of $G$ is a
planar embedding such that the vertices are mapped to points with integer coordinates and the edges
are drawn as paths following the horizontal and vertical grid lines with integer coordinates.
Grid embeddings are especially useful for our geometric reduction as they allow us to construct gadgets with precise geometric properties on an integer grid.
Given a planar graph $G$ on $n$ vertices and a planar representation $P$, Tamassia~\cite{bends} gave an $O(n^2\log n)$-time algorithm that
computes a grid embedding of $G$ that conforms to $P$ --- that is, the faces in the grid embedding correspond exactly to the face descriptions in $P$.
 In addition, his algorithm minimizes the total number of bends
on the edges of the grid embedding. It's worth noting that Tamassia's algorithm produces a grid of polynomial size, which is essential for our reduction to remain polynomial.
In what follows, we assume that we have obtained in polynomial time a grid embedding, $D_\psi$, of $G_\psi$, which serves as the foundation for constructing our geometric gadgets.
See Figure~\ref{fig:grid_embedding} for an example of a formula $\psi$, its graph $G_\psi$, and a grid embedding $D_\psi$ of $G_\psi$.

\section{Reduction to the Euclidean 3-matching Problem}\label{sec:3m}

We construct in polynomial time a set $S_\psi$ of $3m$ grid points, such that $S_\psi$ has
a Euclidean $3$-matching of weight equal to $2m$ if and only if $\psi$ is satisfiable.
A key observation is that
since $S_\psi$ is a set of grid points, the distance between any two of its points is at
least one. Therefore, every $3$-set of $S_\psi$ has weight at least two. Thus, in a $3$-matching
of $S_\psi$ of weight equal to $2m$, every $3$-set must have weight equal to exactly two.
Our reduction strategy maps truth assignments in $\psi$ to optimal 3-matchings in $S_\psi$ by ensuring that the only way to achieve the target weight of $2m$ is to match points in a pattern corresponding to a satisfying assignment.
We now describe the gadgets used in our construction, which include variable gadgets to encode truth values, wire gadgets to propagate these values, and clause gadgets to enforce the SAT constraints.
Due to the complexity of the gadgets, we constantly refer to the figures. 
\begin{figure}[ht]
 	\centering
 	\includegraphics[width=.4\textwidth]{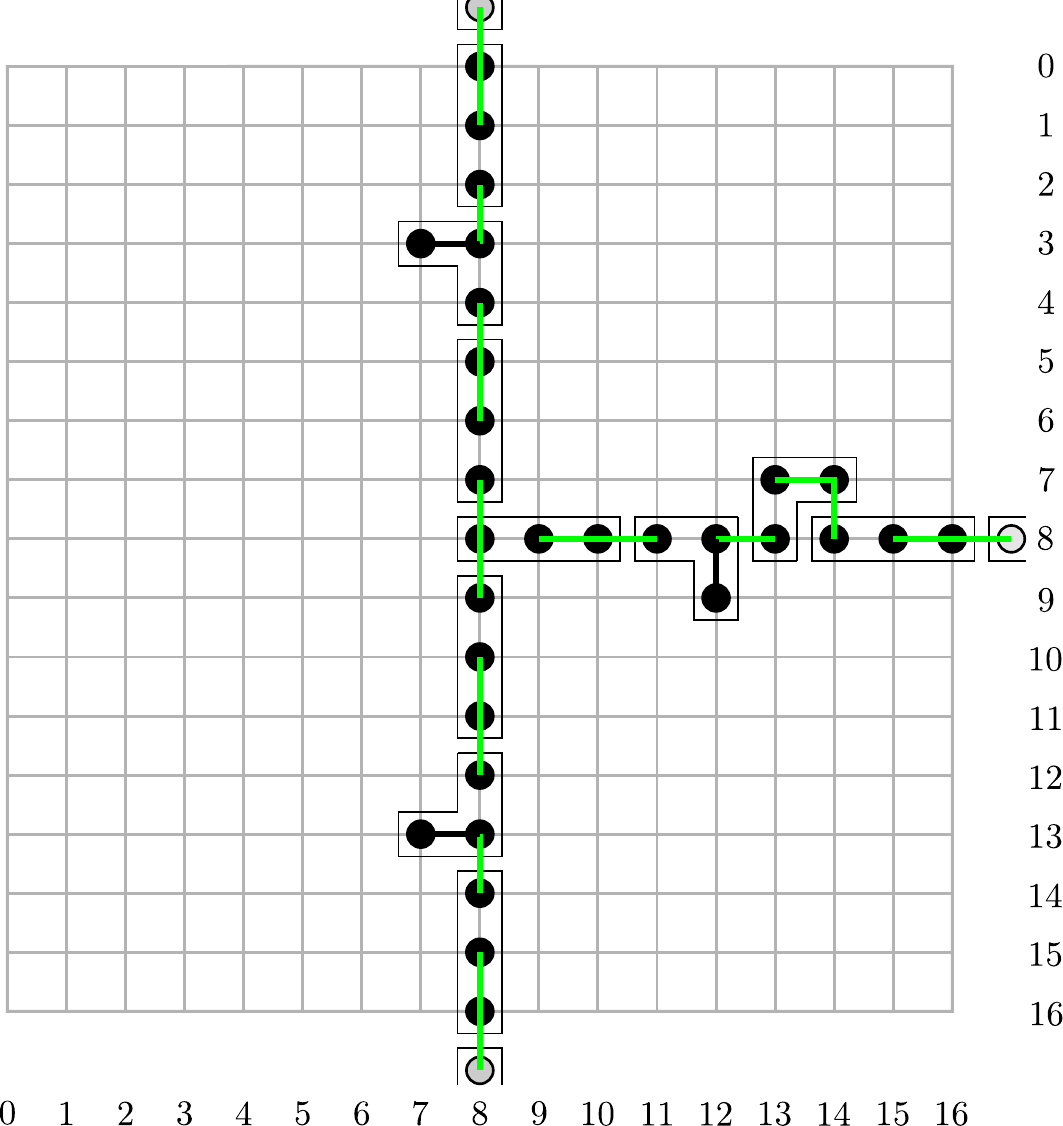}
 	\caption{The variable gadget}\label{fig:var_gadget}
 \end{figure}

The variable gadget, which encodes the TRUE/FALSE state of each boolean variable in our reduction, is shown in Figure~\ref{fig:var_gadget} in a $16 \times 16$ grid.
The points of the gadget are painted black.
The grey points lie outside the gadget. 
The black edges are forced in the sense that their endpoints must
belong to the same $3$-set in any matching with total weight $2m$. 
The $3$-sets of points joined by green paths of length two are the $3$-sets obtained when setting the variable to TRUE; the $3$-sets enclosed by trominoes are the $3$-sets obtained when setting the variable to FALSE. 
A simple case analysis shows that any other $3$-matching different from these two TRUE/FALSE configurations contains a $3$-set of weight strictly greater than two.
We refer to the points by their coordinates. Consider the point at $(8,8)$. 
In a $3$-matching in which every $3$-set has weight equal to two, the point at $(8,8)$ must be matched in one of the following ways:
\begin{itemize}
 \item with $\{(9,8),(10,8)\}$;
 \item with $\{(8,7),(8,9)\}$;
 \item with $\{(8,7),(9,8)\}$;
 \item with $\{(8,9),(9,8)\}$;
 \item with $\{(8,7),(8,6)\}$; or
 \item with $\{(8,9),(8,10)\}$.
\end{itemize}
If $(8,8)$ is matched to $\{(9,8),(10,8)\}$, then the rest of the $3$-sets of the variable gadget are forced and this corresponds to the FALSE configuration shown in Figure~\ref{fig:var_gadget}.
If $(8,8)$ is matched to $\{(8,7),(8,9)\}$, then the rest of the $3$-sets of the variable gadget are forced and this corresponds to the TRUE configuration shown in Figure~\ref{fig:var_gadget}.
If $(8,8)$ is matched to $\{(8,7),(9,8)\}$, then $(10,8)$, $(11,8)$, and $(12,8)$ must be in the same $3$-set; this implies that $(12,9)$
is in a $3$-set of weight greater than two.
By the same argument, $(8,8)$ cannot be matched to $\{(8,9),(9,8)\}$. If $(8,8)$ is matched to $\{(8,7),(8,6)\}$, then $(8,5)$, $(8,4)$ and $(8,3)$ must be in the same $3$-set;
this implies that $(7,3)$ is in a $3$-set of weight greater than two. By a similar argument $(8,8)$ cannot be matched to $\{(8,9),(8,10)\}$.
The variable gadgets propagate their value via \emph{wire gadgets}; these are sequences of rectangular trominoes $t_1,\dots,t_{s}$ enclosing $3$-sets of $S$ such that $t_1$ contains a point adjacent to the variable gadget (i.e., one of the grey points shown in Figure~\ref{fig:var_gadget}).
When the corresponding variable is set to TRUE, this adjacent point is matched to two points inside the variable gadget, which forces a cascading effect where, for all $i = 1,\dots,s-1$, two points inside tromino $t_i$ must be matched with a point of tromino $t_{i+1}$, effectively transmitting the TRUE value along the entire wire.
When the variable is FALSE, the adjacent point is not matched to the variable gadget, allowing the $3$-matching to simply maintain the natural $3$-sets enclosed within each tromino of the wire gadget, thus preserving the FALSE value throughout the wire.
Importantly, these are the only two possible states for a wire that maintain the requirement that all $3$-sets have weight exactly two.

  \begin{figure}[h]
 	\centering
 	\includegraphics[width=1.0 \textwidth]{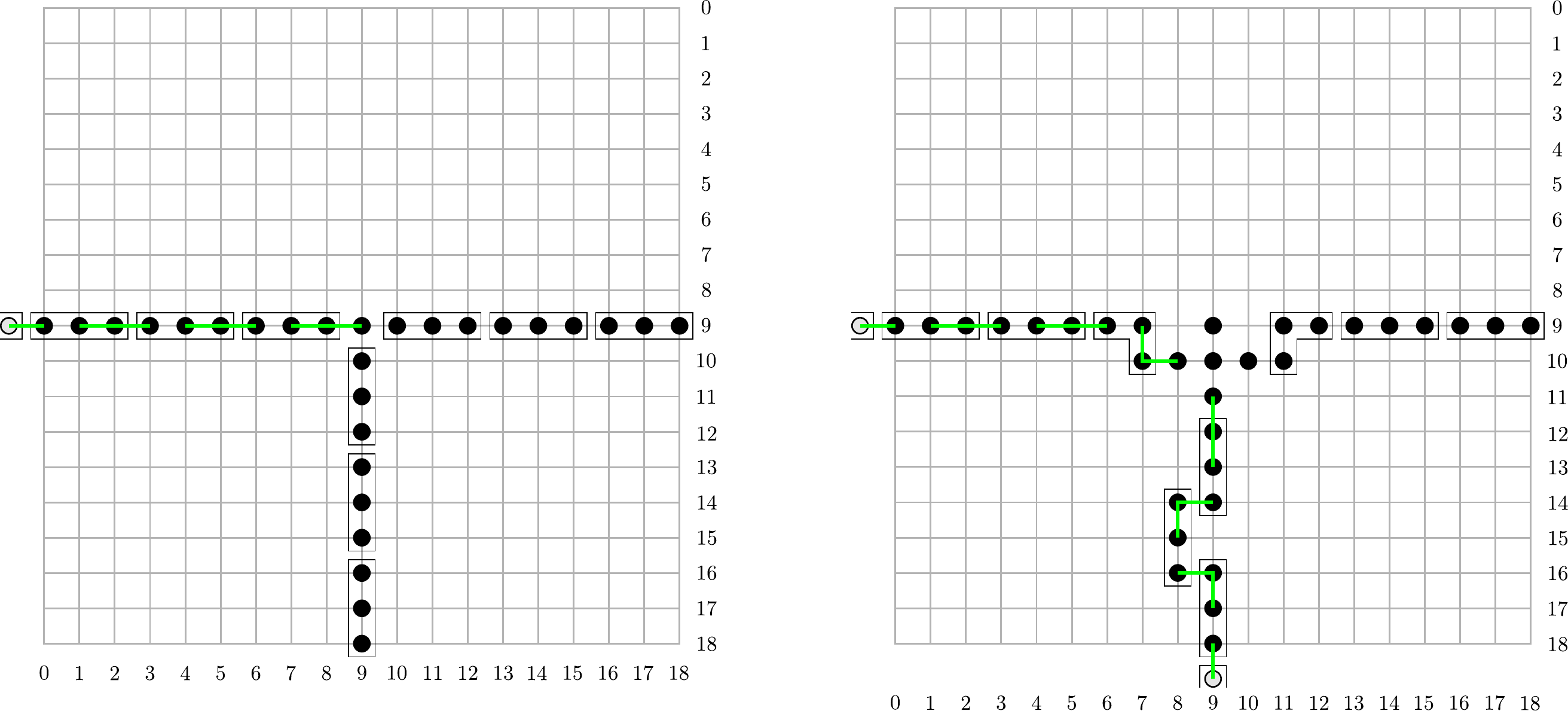}
 	\caption{Clause gadgets}\label{fig:clause_gadget}
 \end{figure}

The clause gadgets, which encode the SAT clauses in our reduction, are shown in Figure~\ref{fig:clause_gadget}: to the left we have the \emph{positive clause gadget}
and to the right the \emph{negative clause gadget}. Each one fits in an $18\times 18$ grid and enforces the clause constraints through careful arrangement of points and trominoes.
\begin{itemize}
    \item In the positive clause gadget there is exactly one point that is not enclosed in a tromino. In order to match this point
     in a $3$-set of weight equal to $2$, one of the wires must carry a TRUE value. Moreover, if any other wire is also carrying a TRUE value, then
     the last tromino of the wire gadget must have two points that cannot be matched to a third point to produce a $3$-set of weight equal to $2$.
     Thus, if there is a $3$-matching of $S$ of weight $2m$, then exactly one of the three wires carries a TRUE value. In Figure~\ref{fig:clause_gadget} (left)
     the green edges depict the wire carrying the TRUE value.

    \item 
    In the negative clause gadget there are exactly five points that are not enclosed in a tromino. For a valid matching of weight $2m$, two of these points must connect with two wires carrying TRUE values, and the remaining three points must form their own $3$-set. This forces the third wire to carry a FALSE value, as any other configuration would leave two points that cannot be matched with the required weight constraint.
    Thus, if there is a $3$-matching of $S$ of weight $2m$, then exactly one of the three wires carries a FALSE value.
    In Figure~\ref{fig:clause_gadget} (right)
     the green edges represent the two wires carrying a TRUE value.
\end{itemize}
These gadget properties ensure that the clause constraints of the original SAT formula are satisfied: positive clauses require exactly one TRUE literal, while negative clauses require exactly one FALSE literal. Thus, a valid $3$-matching of weight $2m$ can exist if and only if there is a satisfying assignment for the original SAT formula. In Figure~\ref{fig:clause_gadget}, the green edges depict valid configurations where the clause constraints are satisfied: in the left figure, exactly one wire carries a TRUE value; in the right figure, exactly one wire carries a FALSE value.

\begin{figure}
 	\centering
 	\includegraphics[width=1.0\textwidth]{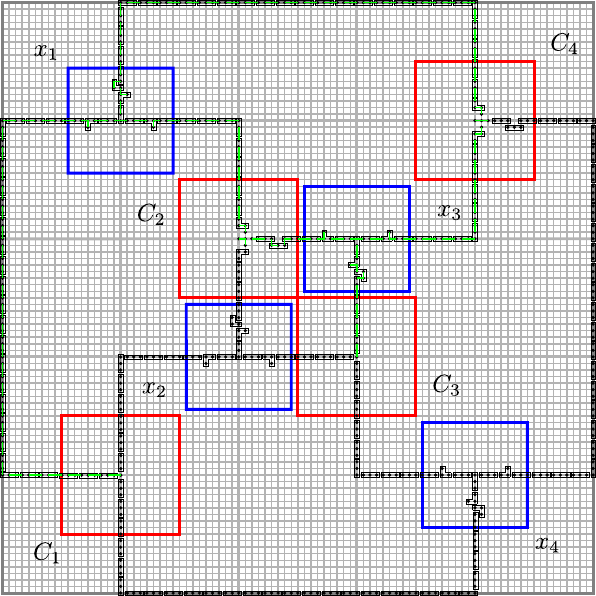}
 	\caption{The grid embedding $D_\psi$ associated to the formula
 	$\psi:=\protect \overbrace{(x_1 \lor x_2 \lor x_4)}^{C_1} \land (  \protect \overbrace{\overline{x_1} \lor \overline{x_2} \lor \overline{x_3})}^{C_2} \land
 	\protect \overbrace{(x_2 \lor x_3 \lor x_4)}^{C_3} \land \protect \overbrace{(\overline{x_1} \lor \overline{x_3} \lor \overline{x_4})}^{C_4}$ }\label{fig:cons}
 \end{figure}

To complete our reduction, we use the grid embedding $D_\psi$ as a template for placing our gadgets. First, we refine the grid by dividing each unit square into an $18 \times 18$ grid of smaller squares, transforming each original edge into a path of $18$ consecutive edges. We replace each variable vertex $x_i$ in $D_\psi$, with a $16 \times 16$ square region,  removing internal edge portions while preserving external connections 
corresponding to the three clauses containing that variable.
Within each such region, we place a variable gadget, rotated to align its three interface points with the preserved edge portions from $D_\psi$.
We similarly replace each clause vertex $C_j$ with an $18 \times 18$ square region containing the appropriate clause gadget. The remaining edge segments become wire gadgets through a systematic partitioning into consecutive triples of points, each enclosed in a rectangular tromino.
Figure~\ref{fig:cons} shows this construction for our sample formula and its grid embedding from Figure~\ref{fig:grid_embedding}. By our gadget properties, a solution to $\psi$ exists if and only if $S_\psi$ has a $3$-matching of weight $2m$. Since we construct
$S_\psi$ in polynomial time, this proves the NP-hardness of the Euclidean $3$-Matching Problem.

\section{The Euclidean $k$-Matching Problem}\label{sec:km}

\begin{figure}[h]
 	\centering
 	\includegraphics[width=.4\textwidth]{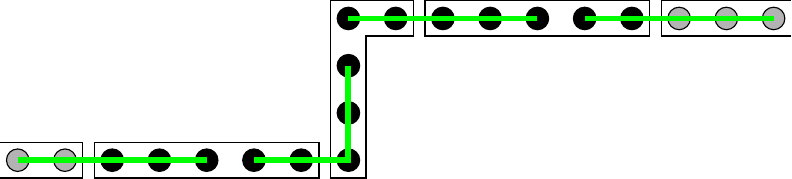}
 	\caption{A wire, for $k=5$ carrying a charge of $3$ from left to right and a charge of $2$ from right to left.}\label{fig:charge}
 \end{figure}

We follow a similar approach as for the case $k=3$. We also have variable and clause gadgets. \emph{Wires}, are now defined as sequences of consecutive $k$-minos.
Wires can now carry more ``charge''. Let $t_1,\dots,t_s$ be a sequence
$k$-minos forming a wire.
Suppose that $i$ of the points at $t_2$ are matched with points of $t_1$; this creates a cascading effect, where for all $j=2,\dots,s$, exactly $i$ points of $t_j$ must be matched with points in $t_{j-1}$ to maintain the minimum weight constraint. In such a case, we say that the wire carries a \emph{charge} of $i$ from $t_1$ toward $t_s$.
Note that, we may consider the charge of the wire going in the opposite direction, so that
with the same notation as the example above, $t_s,t_{s-1},\dots,t_1$ is again a wire, but it carries a charge of $k-i$. See Figure~\ref{fig:charge}.
As in the $k=3$ case, our construction ensures that every $k$-set must have weight exactly $k-1$ in any valid matching, which constrains how points can be matched.

In the context of our charge-based propagation system, recall that for $k=3$, when a variable is set to TRUE,
its corresponding variable gadget outputs a charge of one on each of its three outgoing wires.
For $k > 3$, when a variable is set to TRUE, the corresponding variable gadget distributes charge across three pairs of outgoing wires. Each pair receives a total charge of $k$,
split between a wire carrying $\lfloor k/2 \rfloor$ charge and another wire carrying $\lceil k/2 \rceil$ charge.
These three pairs of wires are connected to the clause gadgets where the variable appears.
The positive clause gadget is designed to ensure that exactly one of its three incoming pairs of wires carries a charge greater than zero (corresponding to exactly one TRUE literal), while the negative clause gadget ensures that exactly two of its three incoming pairs of wires carry a charge greater than zero (corresponding to exactly one FALSE literal).
The implementation of these gadgets requires several building blocks that carefully manipulate charges while maintaining our critical weight constraint of exactly $k-1$ for each $k$-set. We now describe these components in detail.

\subsection*{Variable Gadget}

\begin{figure}[h]
 	\centering
 	\includegraphics[width=.8\textwidth]{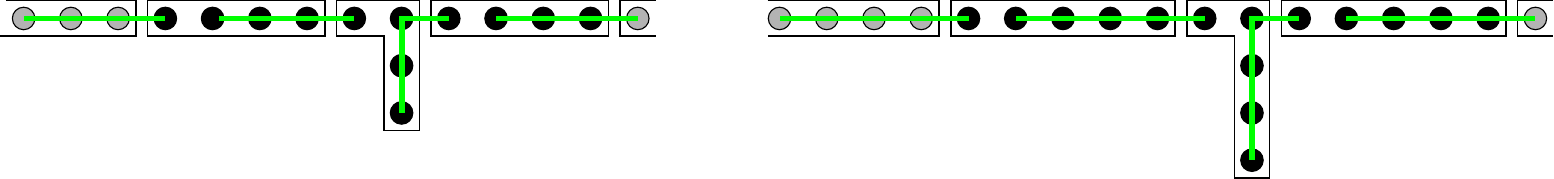}
 	\caption{Fuses for $k=4$ and $k=5$}\label{fig:fuse}
 \end{figure}

A \emph{fuse} is an $L$-shaped $k$-mino, that belongs to the sequence of $k$-minos of a wire that ensures that the charge of the wire does not exceed one. A fuse is constructed by placing two adjacent vertices $v_1$ and $v_2$ with the same $y$-coordinate, followed by a vertical $(k-1)$-mino starting at $v_2$.
Suppose that $t_{i-1}$ and $t_{i+1}$ are the previous and next $k$-minos from the fuse in the wire. 
In this arrangement, $v_1$ is adjacent to the last vertex of the preceding $k$-mino $t_{i-1}$, and $v_2$ is adjacent to the first vertex of the following $k$-mino $t_{i+1}$.
 See Figure~\ref{fig:fuse}.

\begin{figure}[ht]
 	\centering
 	\includegraphics[width=.7\textwidth]{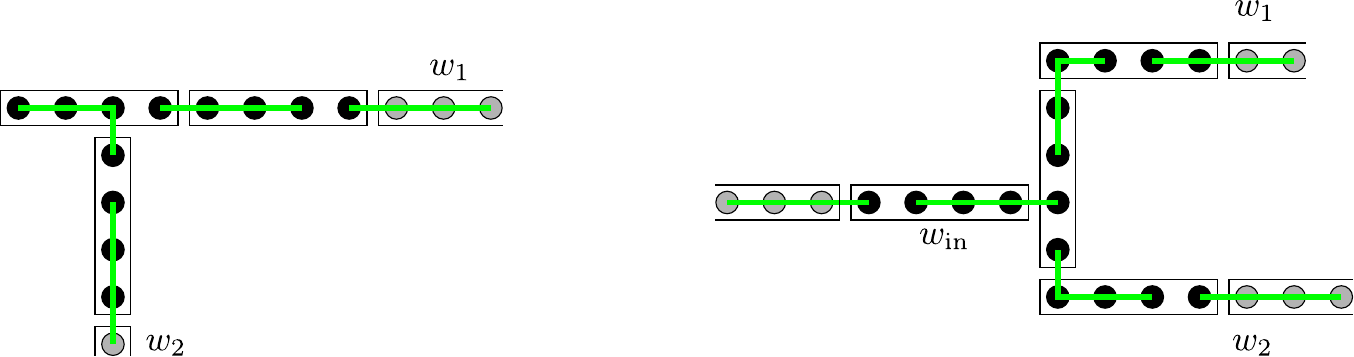}
 	\caption{A switch and an amplifier for $k=4$}\label{fig:switch_amp}
 \end{figure}

A \emph{switch} is a gadget where two wires $w_1$ and $w_2$ begin. The whole construction ensures that either a charge of zero travels through both $w_1$ and $w_2$, or a charge of $k-1$ and a charge of one travels through $w_1$ and $w_2$, respectively. A switch is constructed by placing a horizontal $k$-mino with vertices $v_1,\dots, v_k$ from left to right, then placing a horizontal $k$-mino with its leftmost vertex adjacent to $v_{k}$ (this forms the beginning of wire $w_1$), and finally placing a vertical $k$-mino with its bottommost vertex adjacent to $v_{k-1}$ (this forms the beginning of wire $w_2$). See Figure~\ref{fig:switch_amp}.

An \emph{amplifier} is a gadget that has one incoming wire $w_{\textrm{in}}$ and two outgoing wires $w_1$ and $w_2$. The whole construction ensures that the charge on $w_{\textrm{in}}$ is equal to either zero or one. If the charge at $w_{\textrm{in}}$ is equal to one, then $w_1$ carries a charge of two and $w_2$ carries a charge of $k-1$. An amplifier is constructed by placing a horizontal $k$-mino (this is the last $k$-mino of $w_{\textrm{in}}$), then placing a vertical $k$-mino $t$ whose vertices from top to bottom are $v_1,\dots,v_k$ such that $v_{k-1}$ is adjacent to the last vertex of $w_{\textrm{in}}$, then placing a horizontal $k$-mino with its leftmost vertex adjacent to $v_1$ (this forms the beginning of $w_1$), and finally placing a horizontal $k$-mino with its leftmost vertex adjacent to $v_k$ (this forms the beginning of $w_2$). See Figure~\ref{fig:switch_amp}.

\begin{figure}
 	\centering
 	\includegraphics[width=.8\textwidth]{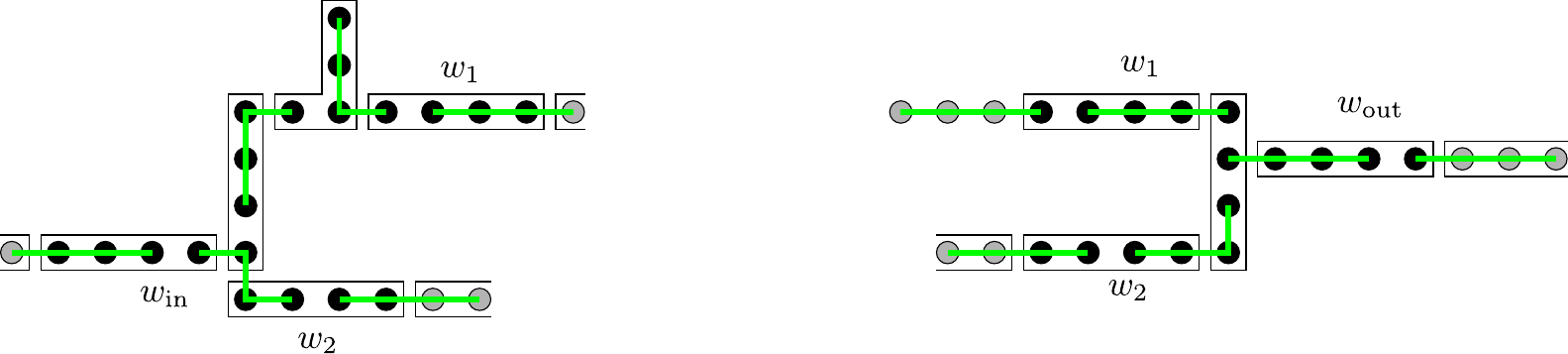}
 	\caption{A splitter and a junction for $k=4$}\label{fig:splitter_junction}
 \end{figure}

A \emph{splitter} is a gadget with an incoming wire $w_{\textrm{in}}$ and two outgoing wires $w_1$ and $w_2$. If $w_{\textrm{in}}$ carries a charge of $s$, then $w_1$ carries a charge of one, and $w_2$ carries a charge of $s-1$. A splitter is constructed as follows. The last $k$-mino of $w_{\textrm{in}}$ is horizontal. Place a vertical $k$-mino whose vertices from top to bottom are $v_1,\dots,v_k$, where $w_{\textrm{in}}$ is adjacent to $v_{k-1}$. Wire $w_1$ is adjacent to $v_1$, and the first $k$-mino of $w_1$ is a fuse to ensure that the charge through $w_1$ does not exceed one. Wire $w_2$ is adjacent to $v_k$. See Figure~\ref{fig:splitter_junction}.

A \emph{junction} is a gadget that has two input wires $w_1$ and $w_2$, and one output wire $w_{\textrm{out}}$.
The whole construction ensures that the incoming
charges are: either zero through both $w_1$ and $w_2$, or one through $w_1$ and some charge $s >1$ through $w_2$. If both $w_1$ and $w_2$ carry zero charge, then $w_{\textrm{out}}$ outputs a charge of zero; if $w_1$ carries charge
equal to one and $w_2$ charge equal to $s$, then $w_{\textrm{out}}$ outputs a charge equal to $s+1$.
A junction is constructed as follows. Place a vertical $k$-mino whose vertices from top to bottom are
$v_1,\dots,v_k$; $w_{\textrm{out}}$ is adjacent to $v_{2}$; $w_1$ and $w_2$ are adjacent to $v_1$
and $v_{k}$, respectively. See Figure~\ref{fig:splitter_junction}.

\begin{figure}[h]
 	\centering
 	\includegraphics[width=0.8\textwidth]{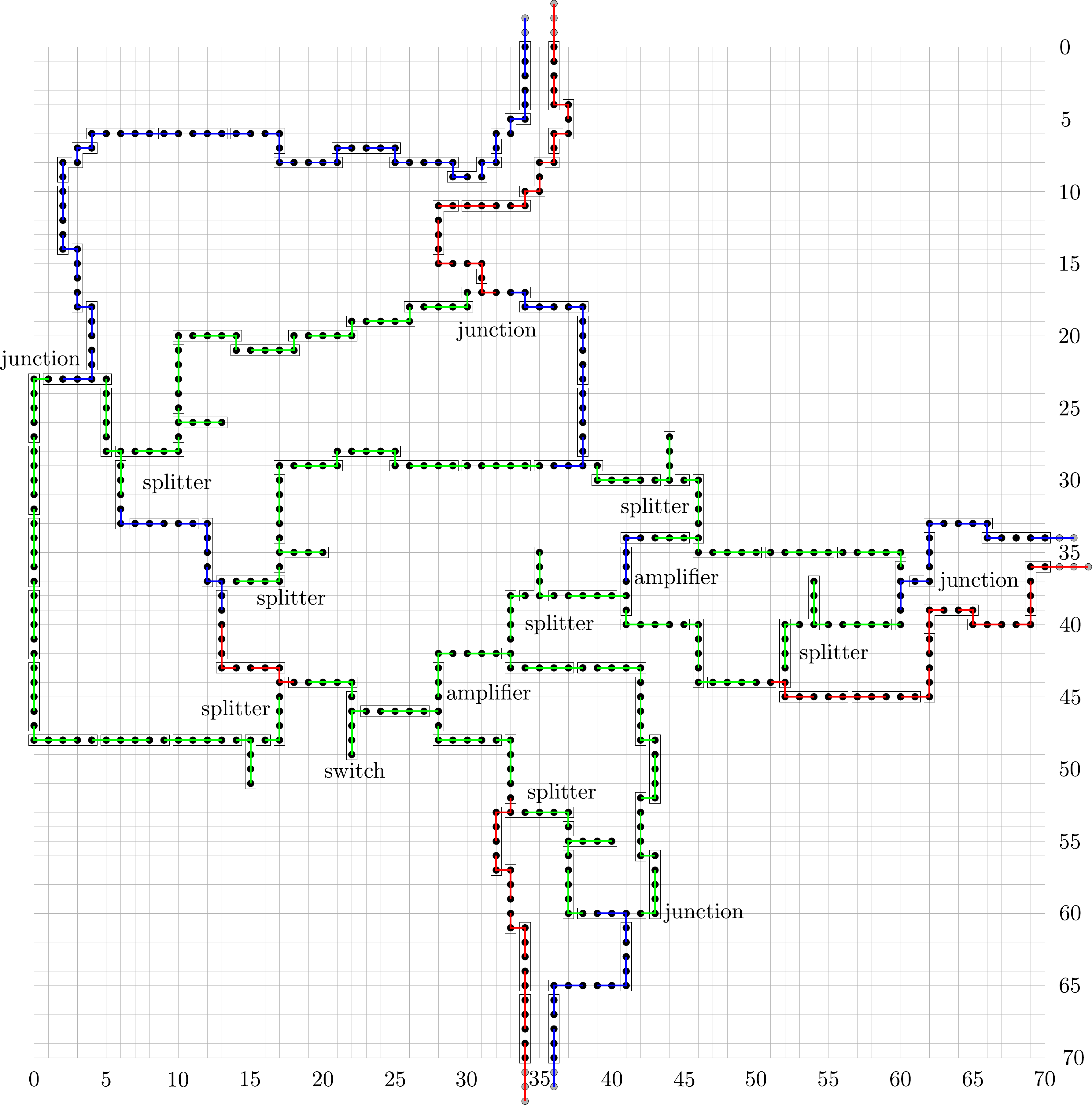}
 	\caption{A variable gadget for $k=5$, the switch turned on, the charges equal to $\lfloor k/2 \rfloor$ are colored blue, the charges equal to $\lceil k/2 \rceil$ are colored red, and all other charges are colored green.}\label{fig:var_k}
 \end{figure}

\begin{figure}[h]
 	\centering
 	\includegraphics[width=0.8\textwidth]{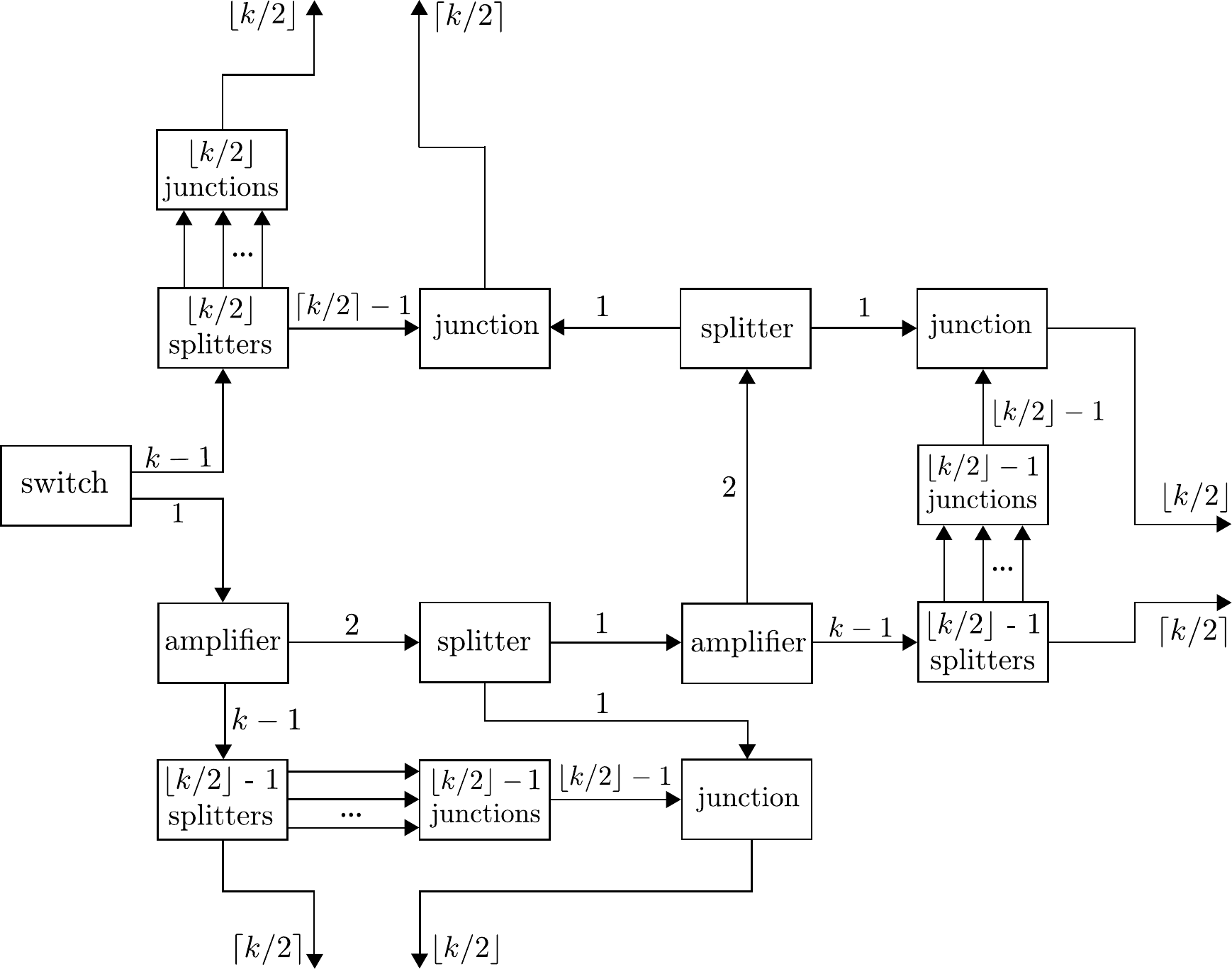}
 	\caption{Variable gadget wiring diagram}\label{fig:var_diagram}
 \end{figure}

We now describe the \emph{variable gadget}. The gadget is enclosed on an $s \times s$ grid, with $s$ even and of polynomial size on $k$. The gadget contains six wires ending at $(s/2-1,0)$, $(s/2+1,0)$, $(s,s/2-1)$, $(s,s/2+1)$,  $(s/2+1,s)$ and $(s/2-1,s)$. Using the previously defined gadgets we ensure that every one of these wires has either charge zero, or
 \begin{itemize}
  \item a charge of $\lfloor k/2 \rfloor$ for the wires ending at  $(s/2-1,0)$,  $(s,s/2-1)$, and $(s/2+1,s)$;
  \item and a charge of  $\lceil k/2 \rceil$ for the wires ending at $(s/2+1,0)$, $(s,s/2+1)$, and $(s/2-1,s)$.
 \end{itemize}
 This can be accomplished in many ways. One such way is as follows.
 We start with a switch in the "on" state (carrying charges of $k-1$ and $1$ on its output wires). We use amplifiers and splitters to convert and distribute this initial charge, ultimately creating a total charge of $3k$ distributed across $3k$ individual wires, each carrying a charge of one. We then use junctions to combine these individual wires, reducing the $3k$ wires to $3$ pairs of wires, where in each pair one wire carries a charge of $\lfloor k/2 \rfloor$ and the other wire carries a charge of $\lceil k/2 \rceil$. We connect these wires to the vertices $(s/2-1,0)$, $(s/2+1,0)$, $(s,s/2-1)$, $(s,s/2+1)$,  $(s/2-1,s)$ and  $(s/2+1,s)$. When the switch is in the "on" state, this corresponds to the variable being set to TRUE, and the six output wires carry the specified charges. When the switch is "off," all output wires carry zero charge, corresponding to the variable being set to FALSE. In Figure~\ref{fig:var_k}, we show a variable gadget for $k=5$ on a $70 \times 70$ grid. In Figure~\ref{fig:var_diagram}, we show a schematic depiction of how the variable gadget is constructed for an arbitrary value of $k$.

\section*{Clause Gadgets}

\begin{figure}[h]
 	\centering
 	\includegraphics[width=0.8\textwidth]{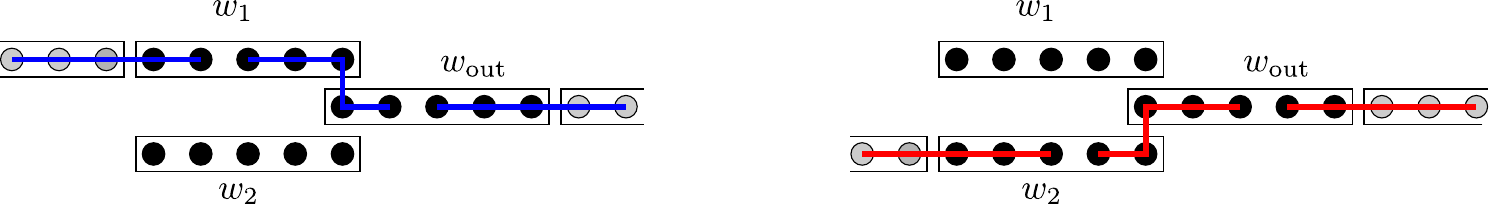}
 	\caption{An XOR-filter for $k=5$,  the two possibilities of exactly one of $w_1$ and $w_2$ carrying a charge distinct from zero are depicted.}\label{fig:xor_filter}
 \end{figure}

Like variable gadgets, clause gadgets are made from smaller building blocks.
 An \emph{XOR-filter} is a gadget that receives two input wires $w_1$ and $w_2$.
 It ensures that either both $w_1$ and $w_2$ carry a charge of zero, or exactly
 one of $w_1$ and $w_2$ carries a non-zero charge (but not both). When exactly one input wire carries a non-zero charge, that charge value is passed through to the output wire $w_{\textrm{out}}$. When both inputs are zero, the output is also zero.
 Moreover, the whole construction ensures that $w_1$ carries either a charge of zero
 or a charge of $\lfloor k/2 \rfloor$, and $w_2$ carries either a charge of zero
 or a charge of $\lceil k/2 \rceil$. 
 An XOR-filter is constructed as follows, the last $k$-minos of both $w_1$ and $w_2$ are horizontal.
 The first $k$-mino $w_{\textrm{out}}$ is also horizontal and its leftmost vertex is adjacent to both $w_1$ and
 $w_2$.
 See Figure~\ref{fig:xor_filter}.

\begin{figure}[h]
 	\centering
 	\includegraphics[width=1\textwidth]{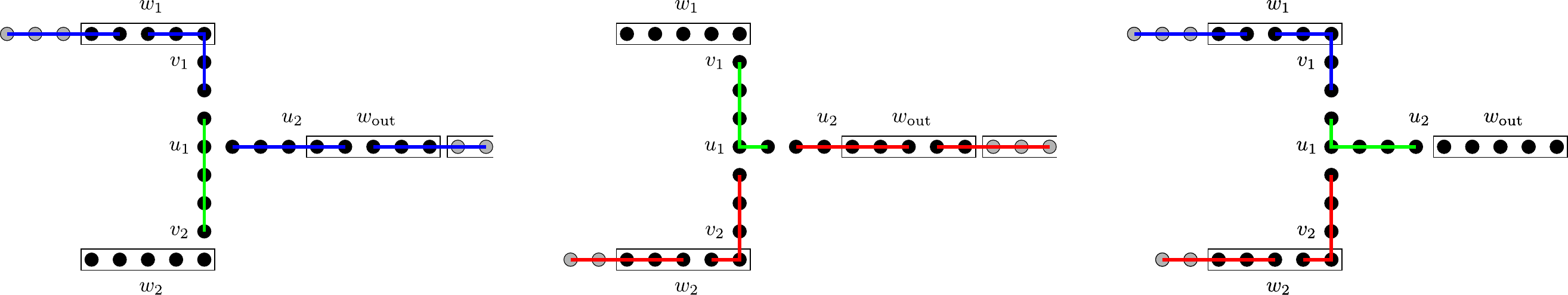}
 	\caption{XOR-enforcers for $k=5$, the three possibilities distinct from both $w_1$ and $w_2$ carrying zero charge are depicted. }\label{fig:xor_gate}
 \end{figure}

 An \emph{XOR-enforcer} is a gadget that receives two input wires $w_1$ and $w_2$. Its construction ensures that at least one of $w_1$ and $w_2$ carries a charge distinct from zero. The whole constructon ensures that $w_1$ carries either a charge of zero or a charge of $\lfloor k/2 \rfloor$, and $w_2$ carries either a charge of zero or a charge of $\lceil k/2 \rceil$. If exactly one of $w_1$ and $w_2$ carries a non-zero charge, that charge is outputted on wire $w_{\textrm{out}}$. If both $w_1$ and $w_2$ carry non-zero charges, then no charge outputted on $w_{\textrm{out}}$. This is achieved as follows. There is a vertical $2 \lceil k/2 \rceil+1$-mino $t_1$. Let $v_1$ be the topmost vertex of $t_1$ and $v_2$ its bottommost vertex. Sorted from top to bottom, let $u_1$ be the $(\lceil k/2 \rceil+1)$-th vertex. There is a horizontal $(2k-2 \lceil k/2 \rceil)$-mino $t_2$, whose leftmost vertex is $u_1$ and whose rightmost vertex is $u_2$. Note that the number of vertices in $t_1 \cup t_2$ is equal to $2k$. Finally, the last vertex of $w_1$ is adjacent to $v_1$, the last vertex of $w_2$ is adjacent to $v_2$, and the first vertex of $w_{\textrm{out}}$ is adjacent to $u_2$. See Figure~\ref{fig:xor_gate}.

\begin{figure}[h]
 	\centering
 	\includegraphics[width=1.0\textwidth]{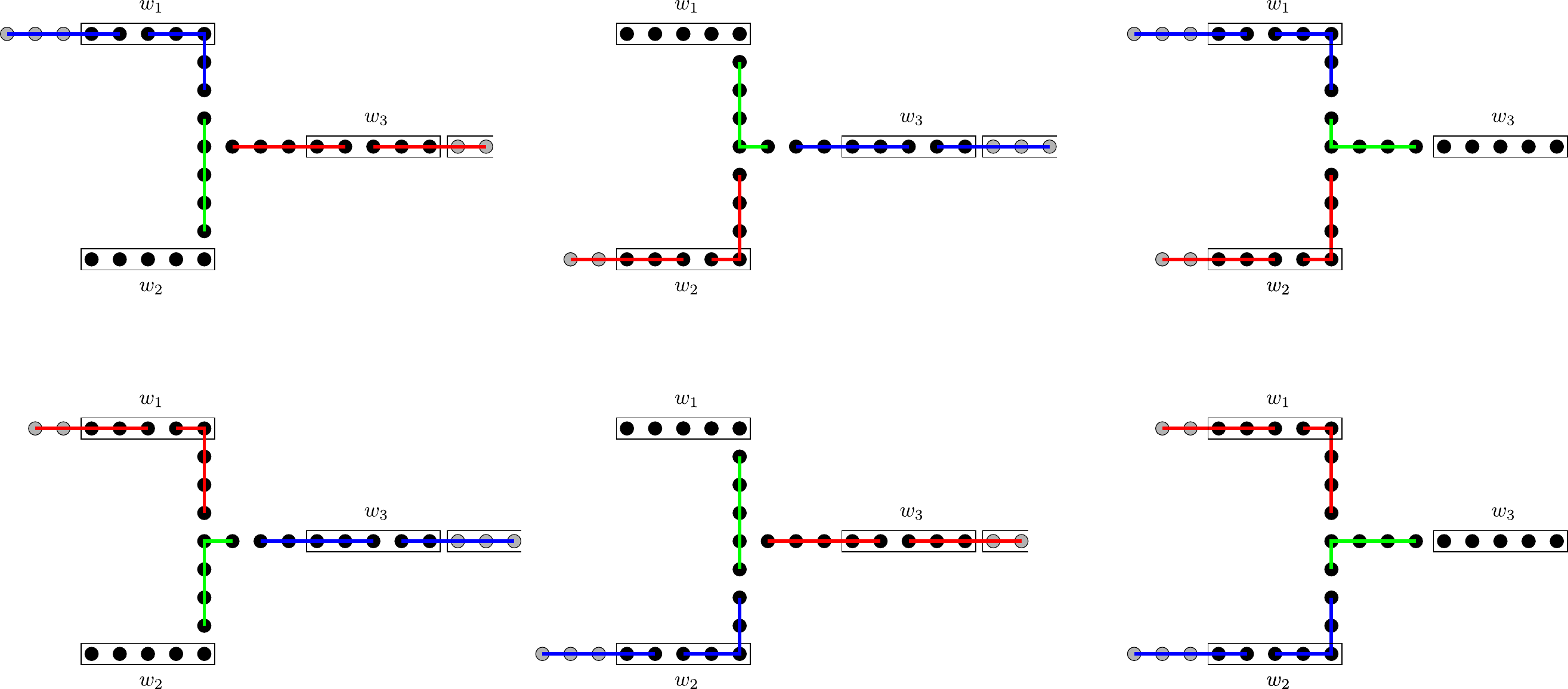}
 	\caption{A $\Delta$-networks for $k=5 $, the six possibilities where exactly two input wires carry a charge distinct from zero are depicted.}\label{fig:delta}
 \end{figure}

 A \emph{$\Delta$-network} is a gadget with three input wires $w_1$, $w_2$ and $w_3$. The whole constructuion ensures that each of these wires carries a charge of zero, $\lfloor k/2 \rfloor$ or $\lceil k/2 \rceil$. The construction ensures that exactly two of these three wires carries a charge distinct from zero; moreover, one of these two wires carries a charge of $\lfloor k/2 \rfloor$ and the other a charge of $\lceil k/2 \rceil$. The gadget is constructed exactly as the XOR-enforcer, where $w_3$ replaces $w_{\textrm{out}}$. See Figure~\ref{fig:delta}.

We now construct the \emph{positive} and \emph{negative} clause gadgets. Both gadgets are enclosed in a $(s-2)\times (s-2)$ grid. In both cases there are six wires $l_1, r_1, l_2, r_2, l_3$ and $r_3$, beginning at  $(0,s/2-2)$, $(0,s/2)$, $(s/2-2,s-2)$, $(s/2,s-2)$,  $(s-2,s/2)$ and $(s-2,s/2-2)$, respectively. For every pair $(l_i,r_i)$ we have that either
 \begin{itemize}
  \item both $l_i$ and $r_i$ have a charge of zero, or
  \item $l_i$ has a charge of $\lfloor k/2 \rfloor$ and $r_i$ has a charge of $\lceil k/2 \rceil$.
 \end{itemize}
 In the positive clause gadget: $l_1$ and $r_3$ are the input wires of an XOR-filter, with output wire $o_1$; $l_2$ and $r_1$ are the input wires of an XOR-filter, with output wire $o_2$; and $l_3$ and $r_2$ are the input wires of an XOR-filter, with output wire $o_3$.
 In the negative clause gadget: $l_1$ and $r_3$ are the input wires of an XOR-enforcer, with output wire $o_1$; $l_2$ and $r_1$ are the input wires of an XOR-enforcer, with output wire $o_2$; and $l_3$ and $r_2$ are the input wires of an XOR-enforcer, with output wire $o_3$.
 Finally, in both cases $o_1, o_2$ and $o_3$ are the input wires of a $\Delta$-network.
 In the case of the positive clause gadget, the XOR-filters ensure that exactly one pair of wires $(l_i,r_i)$ carries a charge greater than zero. In the case of the negative clause gadget, the XOR-enforcers ensure that exactly one pair of wires $(l_i,r_i)$ carries a charge equal to zero.

 \begin{figure}[h]
 	\centering
 	\includegraphics[width=0.8\textwidth]{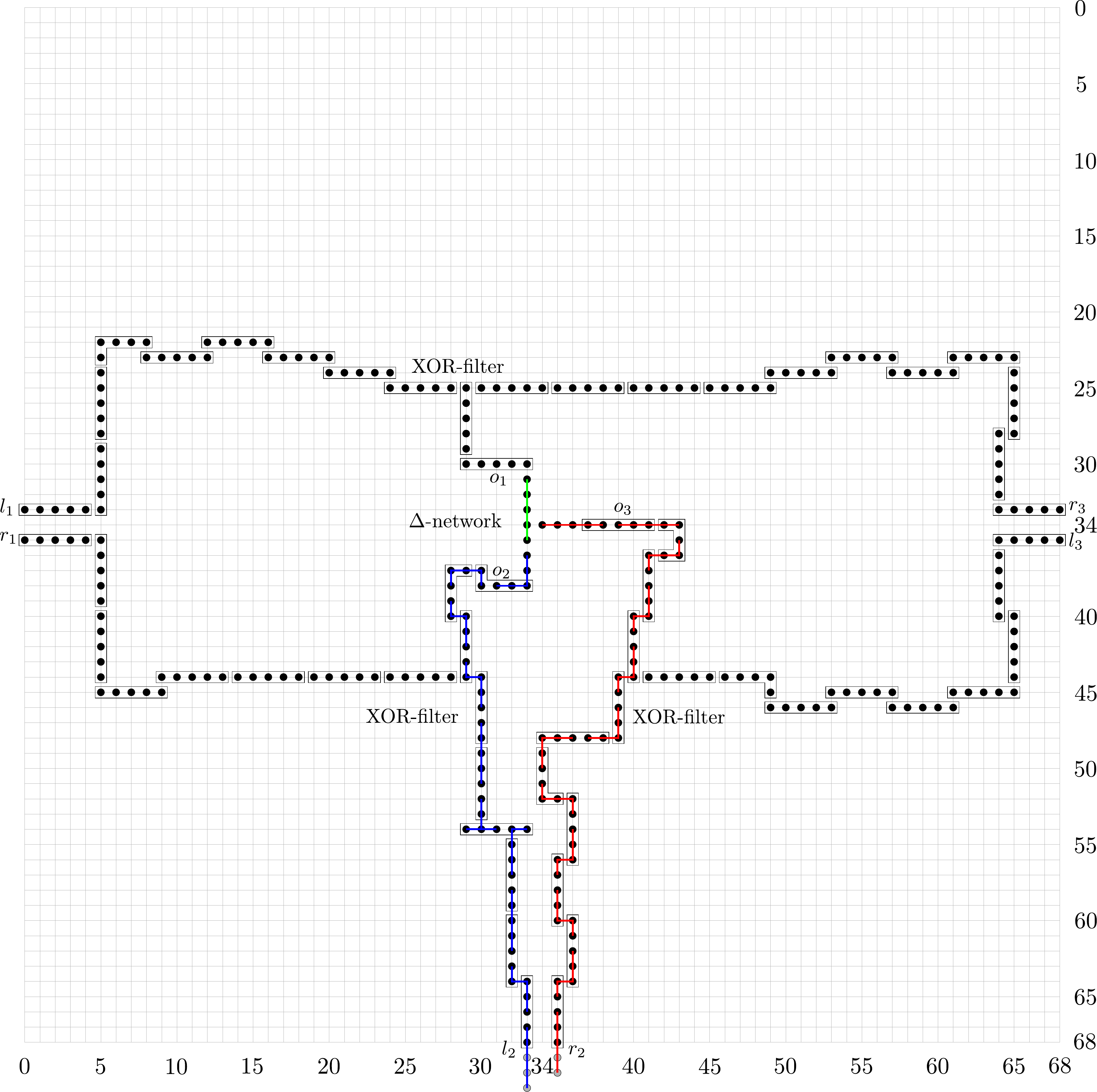}
 	\caption{A positive clause gadget for $k=5$, the charges equal to $\lfloor k/2 \rfloor$ are colored blue, the charges equal to $\lceil k/2 \rceil$ are colored red, and all other charges are colored green.}\label{fig:clause_pos_k}
 \end{figure}

 \begin{figure}[h]
 	\centering
 	\includegraphics[width=0.8\textwidth]{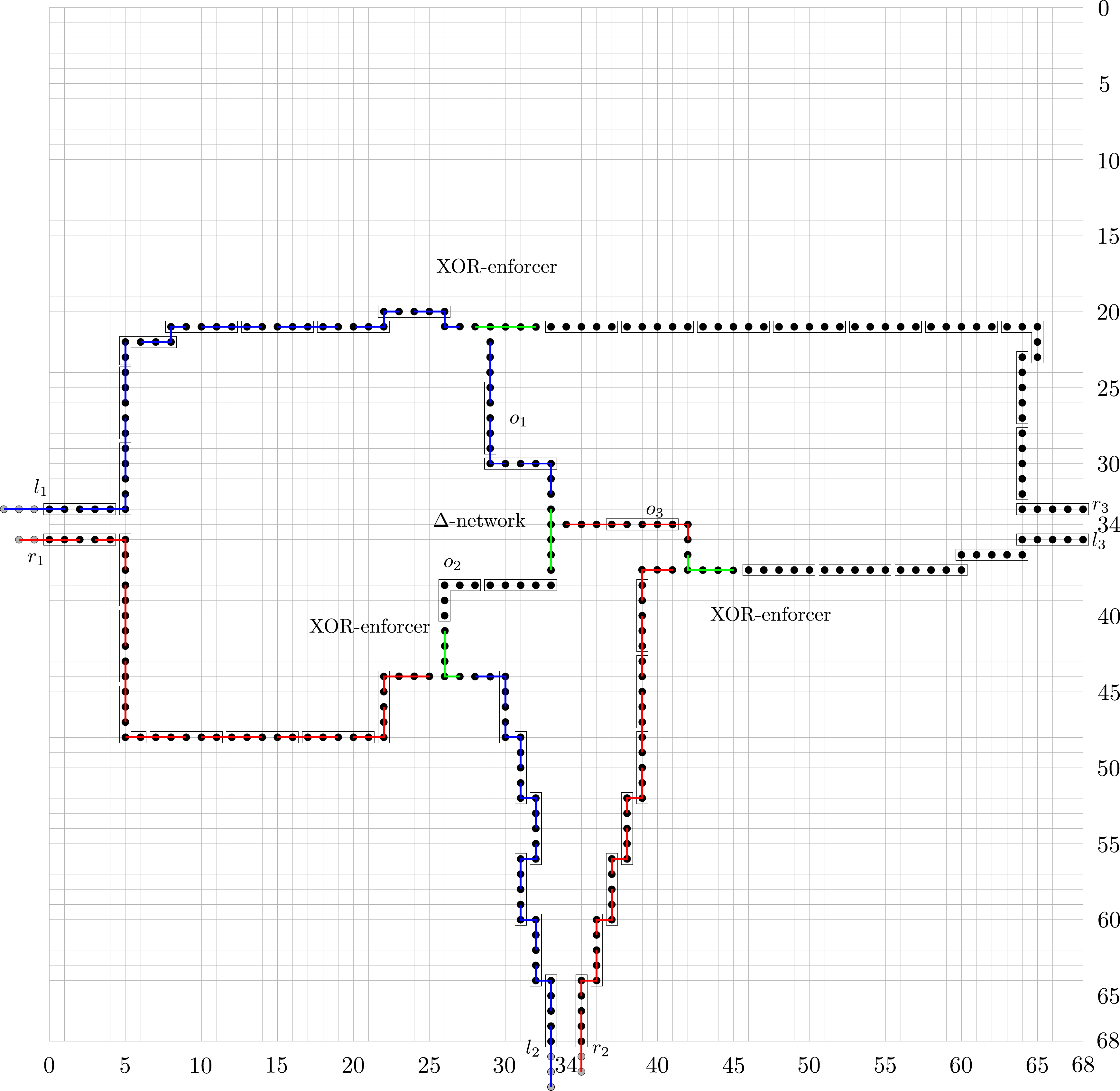}
 	\caption{A negative clause gadget for $k=5$, the charges equal to $\lfloor k/2 \rfloor$ are colored blue, the charges equal to $\lceil k/2 \rceil$ are colored red, and all other charges are colored green.}\label{fig:clause_neg_k}
 \end{figure}

\subsection{Reduction to the Euclidean $k$-matching Problem}

\begin{figure}[h]
 	\centering
 	\includegraphics[width=1.0\textwidth]{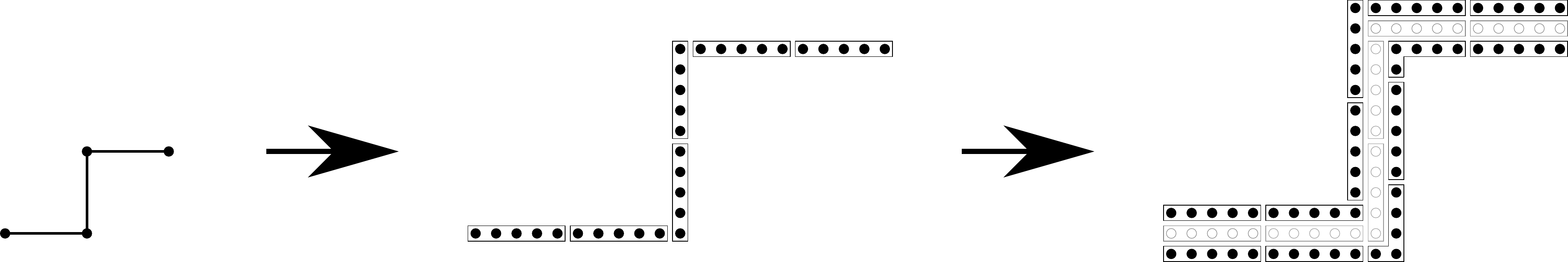}
 	\caption{The edges of $D_\psi$ are first replaced by sequence of $k$-minos, this sequence is then replaced by two parallel wires.}\label{fig:new_wires}
 \end{figure}

 We do a reduction from cubic planar monotone $1$-in-$3$-SAT. As before,
 let $\psi$ be an instance of cubic planar monotone $1$-in-$3$-SAT, and $D_\psi$ be a grid embedding of $G_\psi$. We construct in polynomial time a set $S_\psi$ of $km$ points such that $S_\psi$ has a Euclidean $k$-matching of weight equal to $(k-1)m$ if and only if $\psi$
is satisfiable. As before, every $k$-set of $S_\psi$ has weight at least $k-1$. Therefore, in a $k$-matching of $S_\psi$ of weight equal to $(k-1)m$, every $k$-set must have weight equal
to $k-1$.
As for the case of $k=3$, we use $D_\psi$ as a template. We refine the grid by dividing each unit square into a $ks\times ks$ grid
of smaller squares. Thus every original edge of $D_\psi$ is replaced by a path of $ks$ consecutive edges.
 We replace each variable vertex $x_i$ in $D_\psi$ with an $s \times s$ square region, removing internal edge portions while preserving external connections. Within each such region, we place a variable gadget, rotated to align its three interface points with the preserved edge portions from $D_\psi$.
We similarly replace each clause vertex $C_j$ with an $(s-2) \times (s-2)$ square region containing the appropriate clause gadget.
We orient the edges of $D_\psi$ from the variable vertices to the clause vertices.
Finally, we replace each (old) edge of $D_\psi$ with two wires. One wire is one unit to the left of the edge and the other wire
is one unit to the right. See Figure~\ref{fig:new_wires}.

\clearpage

\section{The Euclidean $k$-path matching Problem}\label{sec:kpaths}

 \begin{figure}[h]
 	\centering
 	\includegraphics[width=0.5\textwidth]{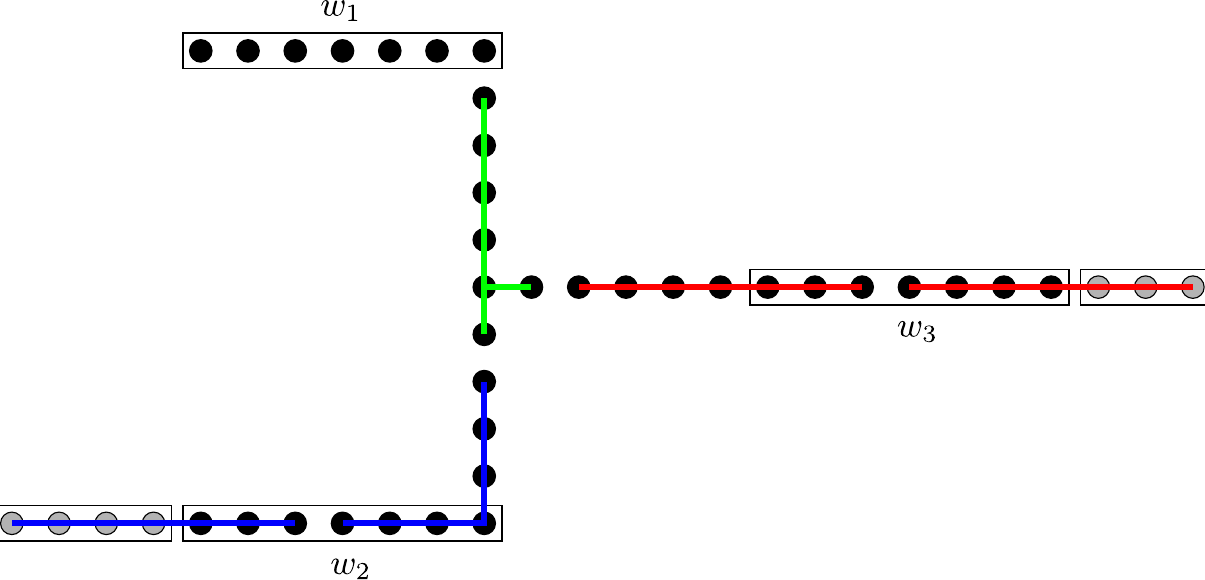}
 	\caption{A $\Delta$-network, for $k=7$, and an instance that produces a non-path tree passing through $k$ points.}\label{fig:delta_t}
 \end{figure}
 
In the Euclidean $k$-matching problem the weight of a $k$-set of points is the cost of its minimum spanning tree. In the case of $k=3$, these trees
are always paths. We now consider the variant where the weight of a $k$-set is the cost of the minimum path joining these points. We call this
variant the Euclidean $k$-matching path problem (E$k$-MPP). It can be verified that for $k\ge 4$, all gadgets in our construction, except for $\Delta$-networks when $k$ is odd,
produce paths. In the case of $k\ge 7$ and odd, a $\Delta$-network may produce a tree that is not a path. See Figure~\ref{fig:delta_t} for an example.

\begin{figure}[h]
 	\centering
 	\includegraphics[width=1.0\textwidth]{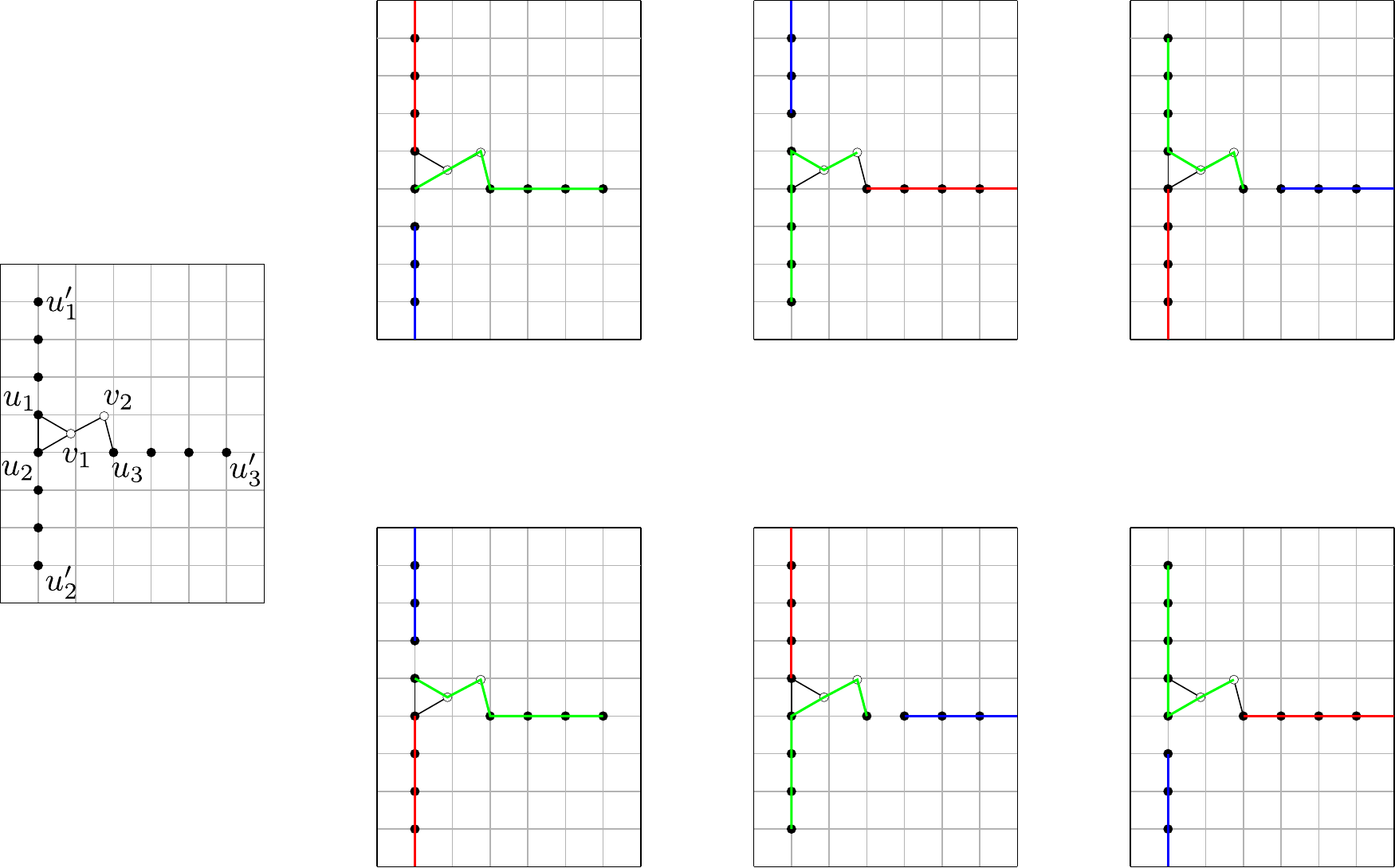}
 	\caption{The $\Delta$-network gadget for E$k$-PP and $k=7$, the white points are \emph{not} grid points; black edges have length equal to one.}\label{fig:delta_path}
 \end{figure}

We show how to modify the $\Delta$-network gadget so that every tree produced is a path. The key idea is to place three points in the vertices of an equilateral triangle
of side length equal to one. This cannot be achieved with all vertices on a grid. We place these three vertices at $u_1:=(0,1)$, $u_2=(0,0)$ and
\[v_1:=\left (\frac{\sqrt{3}}{2}, \frac{1}{2} \right ).\]
We add a vertex 
\[v_2:= \left (\frac{14-3\sqrt{3}+\sqrt{12\sqrt{3}-17}}{20-8\sqrt{3}}, \frac{14-3\sqrt{3}+\sqrt{12\sqrt{3}-17}}{20-8\sqrt{3}}(4-\sqrt{3})-3\right ).\]
Simple arithmetic shows that the distance from $v_1$ to $v_2$ is equal to one. We place: a vertical $(\lceil k/2 \rceil)$-mino $t_1$ that starts
at $u_1$, ends at $u_1'$, and does not contain $u_2$; a vertical $(\lceil k/2 \rceil)$-mino $t_2$ that starts
at $u_2$, ends at $u_2'$, and does not contain $u_1$; and a horizontal $(2k-2\lceil k/2 \rceil-2)$-mino $t_3$ that starts
at $u_3:=(2,0)$, ends at $u_3'$, and does not contain $u_2$. Note that the distance from $v_2$ to $u_3$ is equal to one. 
Finally, the distance between $v_1$ and $u_3$ is equal to 
\[\sqrt{\frac{3-8\sqrt{3}+17}{4}}=1.2393\dots>1.2.\]
The vertices $u_1'$, $u_2$, and $u_3'$ are adjacent to the wires $w_1$, $w_2$, and $w_3'$, respectively. This ends the description of the
new $\Delta$-network gadget. See Figure~\ref{fig:delta_path}.

The reduction follows exactly as for the Euclidean 
$k$-matching problem, where given an instance $\psi$ on $n$ variables of cubic planar monotone $1$-in-$3$-SAT, we construct
an instance $S_\psi$ of E$k$-MPP on $km$ points. We have the following caveat. The coordinates of the non-grid vertices 
in the construction are given with decimal expansions of $t$ decimal places.

We now choose an appropriate value of $t$. Let $\epsilon=\epsilon(t) >0$, such that the lengths of each of the four edges $v_1u_1, v_1u_2, v_1v_2, v_2u_3$
in the $\Delta$-network gadgets are in the interval
\[[1-\epsilon,1+\epsilon].\]
Since $\psi$ has $n$ variables, it also has $n$ clauses, and thus there are $n$ $\Delta$-network gadgets in our construction. 
We choose $t$ sufficiently large so that \[4n\epsilon<0.2.\]
It can be verified that $t$ can be chosen so that it is $O(\log n)$, and thus polynomial in $n$. We now have that $\psi$ is satisfiable if and only if $S_\psi$ has a $k$-matching
with paths whose sum of weights is at most
\[(k-1)m+0.2. \]

\section{Conclusions}

We have proved that the 3-Matching Problem remains NP-hard in its Euclidean version, resolving a long-standing open question in the literature. Moreover, our reduction technique from Cubic Planar Monotone 1-in-3 SAT is shown to be extensible to the Euclidean $k$-Matching Problem for every $k>3$. 
The reduction for $k>3$ requires the use of new building blocks to construct the necessary gadgets. Finally, a small modification of one of these gadgets in the reduction for the E$k$-MP allows us to prove NP-hardness of a variant, E$k$-PMP, in which the weight of the subsets is given by the length of the shortest path visiting all points.
 
The development of approximation algorithms remains a promising research direction. Despite the application of various heuristic techniques to the general 3-Matching Problem—including branch-and-bound, greedy algorithms, local search, simulated annealing, and genetic algorithms—no approximation algorithms with provable guarantees are currently known. We believe that exploiting the geometric properties inherent in the Euclidean version could yield crucial insights for developing such algorithms.
E$k$-PMP is closer to TSP (path connectivity matters), whereas E$k$-MP resembles clustering. Thus, the problems may admit different approximation algorithms.

\section*{Acknowledgments}
This work began during the 4th Reunion of Optimization, Mathematics, and Algorithms Workshop \-(\emph{ROMA2023}). We thank the participants for fruitful scientific discussions.
José-Miguel Díaz-Báñez, José-Manuel Higes and Miguel-Angel Pérez-Cutiño were partially supported by grants PID2020-114154RB-I00 and TED2021-129182B-19
I00 funded by MCIN/AEI/10.13039/501100011033 and the European Union
NextGenerationEU/\-PRTR. Ruy Fabila-Monroy was partially supported by 
CONACYT-FORDECYT-PRO\-NA\-CES/39570/2020.
Nestaly~Mar\'in was partially supported by UNAM Postdoctoral Program (POSDOC). Miguel-Angel Pérez-Cutiño was partially supported by grant DIN2020-011317 funded by MCIN/AEI/10.13039\-/501100011033 and the European Union
NextGenerationEU/PRTR.
Pablo Pérez-Lantero was supported by project DICYT 042332PL Vicerrector\'ia de Investigaci\'on, Desarrollo e Innovaci\'on USACH (Chile).

\bibliography{references}
\bibliographystyle{abbrv}

\end{document}